\documentstyle[12pt]{article}
\begin{document}
\def\lsim{\:\raisebox{-0.5ex}{$\stackrel{\textstyle<}{\sim}$}\:}
\def\gsim{\:\raisebox{-0.5ex}{$\stackrel{\textstyle>}{\sim}$}\:}
\bigskip
\bigskip
\begin{center}
\large {\bf Neutrinos at high energy accelerators}\footnote{Lectures given at
the 11th Mt. Sorak Symposium (July 20-25, 1992), Sok-cho, Korea} \\
\bigskip
\bigskip
Probir Roy \\
Tata Institute Of Fundamental Research \\
e-mail: probir@tifrvax.bitnet \\
\bigskip
\bigskip
\end{center}
\bigskip

\noindent $\bullet$ PREAMBLE

\bigskip

\noindent $\bullet$ BRIEF HISTORY AND PRELIMINARIES

\bigskip

\noindent $\bullet$ QUICK REVIEW OF BASIC NEUTRINO PROPERTIES

\bigskip

\noindent $\bullet$ CHARGED CURRENT NEUTRINO PROCESSES

\bigskip

\noindent $\bullet$ NEUTRAL CURRENT NEUTRINO PROCESSES

\bigskip

\noindent $\bullet$ VERY HEAVY NEUTRINOS

\bigskip

\noindent $\bullet$ CONCLUDING SUMMARY

\newpage

\noindent $\bullet$ PREAMBLE

\bigskip

Why is neutrino physics interesting today?  I can list four reasons.
First, our neutrino detectors actually disprove a claim made by the
inventor of the neutrino more than sixty years ago.  In his 1930 letter
to the participants at the Radioactivity Conference in T\"ubingen, Pauli
had proposed an undetectable neutral particle.  Second, neutrinos --
either as beams or as final state products -- constitute clean and
accurate probes into leptonic and semileptonic weak processes since they
appear to have no interactions other than the weak one.  Third, neutrinos
provide a window to catch a glimpse of various types of possible new
physics beyond the Standard Model -- seesaw mechanisms, new generations,
left-right symmetries, grand unified theories etc.  Finally, the study of
primary cosmic neutrinos, i.e. neutrino astronomy, is beginning to unravel
mysteries of various astrophysical objects.

A major part of neutrino physics is concerned with relatively low-energy
neutrinos or antineutrinos.  These can come from astronomical sources such
as the sun, a supernova, etc.  Alternatively, they can be produced from
weakly decaying terrestrial sources that are at rest or moving slowly.  In
these lectures we shall not cover those aspects.  Rather, our focus will
be on ultra-GeV neutrinos.  Such neutrinos can and do get produced by
cosmic ray interactions in the earth's atmosphere but their fluxes are low
except below 1 GeV or so.  So we shall confine ourselves to ultra-GeV
neutrinos produced at high energy accelerators from the decays of
fast-moving mesons.  There are many machines
where good quality beams of such neutrinos have been used or are available.
Quite rich physics has emerged out of scattering experiments with such
beams.  We shall review the essentials of this physics and show in a
selective way how some
of these have become cornerstones of the Standard Model today.
Additionally, we shall try to go a little bit beyond the SM and speculate
on the existence of new very heavy neutrinos ($\gsim$ tens of GeV in mass)
and discuss possible production mechanisms and search strategies for them.
\bigskip
\bigskip

\noindent $\bullet$ BRIEF HISTORY AND PRELIMINARIES \\

\nobreak
Till the beginning of the sixties most neutrinos, studied directly or
indirectly in laboratories, were products from nuclear $\beta$-decay $(Z
\mp 1,A) \rightarrow (Z,A) ~e^\mp \displaystyle \left(\matrix{\bar \nu_e
\cr \nu_e}\right)$ or from muon decay $\mu^\mp \rightarrow e^\mp
\displaystyle \left(\matrix{\bar \nu_e \nu_\mu \cr \nu_e \bar
\nu_\mu}\right)$.  An important example was the Cowan-Reines experiment
[1] which used electron antineutrinos from $\beta$-decaying neutron-rich
fragments produced from the fission of $^{235}U$ in a reactor to study the
reaction $\bar\nu_e + p \rightarrow n + e^+$.

In modern times ultra-GeV neutrino beams are generated in proton (or
antiproton) accelerators when the primary beam hits a target producing
copious numbers of pions and kaons decaying as $\pi \rightarrow
\displaystyle \left(\matrix{\mu\bar\nu_\mu \cr \bar\mu \nu_\mu}\right)$ or
$K \rightarrow \displaystyle \left(\matrix{\mu\bar\nu_\mu \cr \bar\mu
\nu_\mu}\right)$.  Thus these are largely muon neutrinos or antineutrinos.
 A separation can be made between the two by sweeping out the positively
and negatively charged mesons in different directions.  An illustrative
sketch of how a secondary neutrino beam is obtained is shown in Fig. 1.

\vspace{2.5in}

For a primary proton beam, the positives among the forward going mesons
outnumber the negatives.  In fact, at Fermilab, the ratio $n(\pi^+) ~:~ $
$n(\pi^-)$ among such hadrons is typically $\sim 10:1$.  This means that
such machines yield more intense $\nu_\mu$ beams than $\bar\nu_\mu$ ones.
Fig. 2 shows [2] how $\bar\nu_\mu$ beams at various accelerators have
become progressively more intense from 1960 to 1990.  It is also possible
to have an electron neutrino or antineutrino beam by exploiting $K_{e3}$
decay $K \rightarrow \pi\displaystyle\left(\matrix{e\bar\nu_e \cr \bar e
\nu_e}\right)$, but this is a harder process.

\newpage

\vspace*{4in}

The deduction of the energy spectrum of a secondary beam of neutrinos or
antineutrinos is a complicated and tricky procedure.  It consists
basically of two major steps : (1) careful monitoring of the momenta of
the charged pions and kaons produced and their two body leptonic decay
modes  (indeed, depending on whether the
permitted range of $\pi/K$ momenta is wide or narrow, one would get a
wideband or narrowband neutrino beam) and (2) direct measurement of the
charged current induced quasielastic processes $\nu_\mu n \rightarrow
\mu^- p$, $\bar\nu_\mu p \rightarrow \mu^+ n$ from a nuclear target.
Typical examples of $\nu,\bar\nu$ spectra, obtained in the early days of
Fermilab, are shown in a theoretically idealized form in Fig. 3.  Though
the Monte-Carlo derivation of the neutrino spectrum has become a reliable
technique these days, some uncertainties do persist and contribute
systematic errors to absolute cross section measurements.  These are,
however, absent from flux-independent ratios of cross sections.

\newpage

\vspace*{2in}

One of the first spectacular experiments with O(GeV) muon neutrinos was
done by Lederman et al
who concluded, from the failure to observe the reaction $\nu_\mu n
\rightarrow
ep$, that $\nu_\mu /\!\!\!\!\!= \nu_e$.  Since cross sections for such
quasi-elastic processes are small, in the early seventies experimentalists
started studying deep inelastic scattering processes
$$
(\nu_\mu,\bar\nu_\mu)N \rightarrow \mu^\pm X
$$
with very large targets (e.g. huge heavy liquid bubble chambers such as
GARGAMELLE).  These yielded sizable rates and eventually led to the
discovery of neutral current induced processes
$$
(\nu_\mu,\bar\nu_\mu)N \rightarrow (\nu_\mu,\bar\nu_\mu)X.
$$
Since then there have been many subsequent efforts directed towards the
study of
these two types of processes.  In fact, it is fair to say that these
provide the reference frame for neutrino physics at high energy
accelerators these days.  (Of course, purely leptonic scattering
processes, e.g. $\nu_\mu e \rightarrow \nu_\mu e,\mu\nu_e$ and
$\bar\nu_\mu e \rightarrow \bar\nu_\mu e,\bar\mu \nu_e$ and more exotic
reactions such as $(\nu_\mu,\bar\nu_\mu)N \rightarrow {\rm (dimuons)} X$ have
also been studied).
Based on them, finer studies have been made on the excitation of heavy
flavours, Quantum Chromodynamics tests, searches for electroweak physics
beyond the Standard Model, etc.  These experiments have always been done
with large ($> 10^3$ cubic metres) and heavy ($> 10^2$ tons) detectors.
After the extinction of bubble chambers, these detectors have largely been
hadron calorimeters laced with layers of counters.  A list of some of the
recent and current detectors is given in Table 1.  (N.B. this is not a
comprehensive list).

\newpage

\textwidth 6.5in
\begin{center}
\begin{small}
\begin{tabular}{|l|p{1.4in}|l|p{2in}|l|}
\hline
\multicolumn{1}{|c|}{Name} &
\multicolumn{1}{c|}{Collaboration}&
\multicolumn{1}{c|}{Accelerator}&
\multicolumn{1}{c|}{Description}&
\multicolumn{1}{c|}{Status}\\
\hline
CDHS & CERN--Dortmund--Heidelberg--Saclay&
SPS,CERN &
Magnetized iron toroids instrumented with scintillator strips and drift
chambers
& Finished \\
&&&&\\
CHARM 1 & CERN--Hamburg--Amsterdam--Rome--Moscow & SPS,CERN &
Marble plates interspersed with scintillation counters and drift tubes &
Finished\\
&&&&\\
US/JAP & United States--Japan $\nu_\mu e,\bar\nu_\mu e$ expt. & BNL &
170 ton detector with liquid scintillators and proportional drift tubes &
Finished\\
&&&&\\
FMMF & Fermilab--MIT--Michigan--Florida State & TEVATRON &
200 ton calorimeter finegrained with flash chambers and proportional tubes
& Finished\\
&&&&\\
CCFR & Columbia--Chicago--Fermilab--Rochester & TEVATRON &
$\sim$ 700 ton hadron calorimeter of steel plates interlaced with spark
chambers and liquid scintillators; for neutrino and antineutrino
scattering from nucleons only &
Running\\
&&&&\\
CHARM II & CERN--Hamburg--Amsterdam--Rome--Moscow & SPS,CERN &
$\sim$ 690 ton detector of thick glass plates with scintillation counters
and streamer tubes; can study neutrino and antineutrino scattering from
both electrons and nucleons & Running\\
&&&&\\
\hline
\end{tabular}
\end{small}
\end{center}
\begin{center}
{\bf Table 1. Some recent and current neutrino detectors}\\
\end{center}
\bigskip
\bigskip
\noindent $\bullet$ QUICK REVIEW OF BASIC NEUTRINO PROPERTIES

\bigskip
\nobreak

\noindent {\it Neutrinos in the SM}

\bigskip
\nobreak
In the SM the left-chiral neutrino $\nu_{\ell L} ~(\ell = e,\mu,\tau)$ appears
in the same flavour doublet as its charged counterpart $\ell^-_L$.
However, while the right-chiral $\ell^-_R$ does make an entry as a flavour
singlet, $\nu_{\ell R}$ does not.  Thus the neutrino $\nu_\ell$ cannot
have a Dirac mass owing to the latter's absence.  On the other hand, any
Majorana mass term for it would be lepton number violating and the SM has
lepton conservation built into it.  Thus the three neutrinos
$\nu_e,\nu_\mu$ and $\nu_\tau$ are massless in this model.  The
experimental upper limits on their Dirac masses are 7.3 eV, 0.27 MeV and
35 MeV respectively [3].  Only in scenarios going beyond the SM, e.g.
left-right symmetric or grand unified theories [4], do the right-chiral
components $\nu_{eR},\nu_{\mu R}$ and $\nu_{\tau R}$ occur.

Unlike other fermions of the SM, the neutrinos here have only gauge
interactions and no Yukawa couplings.  The interaction terms in the
Lagrangian for the charged current (CC) and neutral current (NC) vertices
can be written respectively as
$$
{\cal L}^{CC}_I = - {g \over \sqrt{2}} \sum_\ell \left[{1\over2}\bar\ell
\gamma^\mu (1 - \gamma_5) \nu_\ell W^-_\mu +
h.c.\right]~~~~~~~~~~~~~~~~~~~~~ ~~~~~~~~~~~~~~~~~~~~~~~~~~~~~~~~~~~~~~~~
\eqno (1)
$$
and
$$
{\cal L}^{NC}_I = - {g \over 2 \cos \theta_W} \sum_\ell
{1\over2} \bar\nu_\ell \gamma^\mu (1 - \gamma_5)\nu_\ell Z_\mu.
{}~~~~~~~~~~~~~~~~~~~~~~~~~~~~ ~~~~~~~~~~~~~~~~~~~~~~~~~~~~~~~~~~~~~~~~~
\eqno (2)
$$
Here the semiweak coupling $g$ and the positron charge $e$ are related
through the Weinberg angle $\theta_W$ by $g = e/\sin \theta_W$, while $g$
is given in terms of the Fermi constant $G_F = 1.166365(16) \times
10^{-5}$ GeV$^{-2}$ via $g^2/(8M^2_W) = G_F/\sqrt{2}$.  Furthermore, $x_W
\equiv \sin^2 \theta_W = 1 - M^2_W/M^2_Z$.  It is noteworthy that in the
SM all neutrino currents have the $V-A$ form.  Moreover, on account of the
masslessness of the neutrinos, there is no mixing among them.

\bigskip

\noindent {\it Illustrations of CC and NC phenomena}

\bigskip
\nobreak
The leptonic decay of the $W$-boson is the simplest charged current
process involving the tree-level vertex (1).  One has the partial width
$$
\Gamma (W^- \rightarrow \ell^- \bar\nu_\ell) = {G_F \over \sqrt{2}} {M^3_W
\over 6\pi} \simeq 214 ~{\rm MeV}.
\eqno (3a)
$$
The total leptonic width of the $W$ is
$$
\Gamma^{\rm lep.}_W \equiv \sum_\ell \Gamma(W^- \rightarrow \ell^-
\bar\nu_\ell) \simeq 642 ~{\rm MeV}.
\eqno (3b)
$$
Similarly, the invisible decay mode of the $Z$ into a specific
neutrino-antineutrino channel is the archetypal neutral current
process -- employing the vertex (2):
$$
\Gamma(Z \rightarrow \nu_\ell \bar\nu_\ell) = {G_F \over \sqrt{2}} {M^3_Z
\over 12\pi} \simeq 155 ~{\rm MeV}.
\eqno (4a)
$$
Consequently, the total invisible $Z$-width is
$$
\Gamma^{\rm Inv.}_Z \equiv \sum_\ell \Gamma(Z \rightarrow \nu_\ell
\bar\nu_\ell) \simeq 465 ~{\rm MeV}.
\eqno (4b)
$$
Defining $\Gamma^{\rm Inv.}_Z \equiv N_\nu \Gamma(Z \rightarrow \nu_\ell
\bar\nu_\ell)$, one finds [4] from the LEP 1 data that $N_\nu = 2.99 \pm
0.04$.  Of
course, one could have one or more generations [5] with neutrinos that
have masses $> {1\over2} M_Z$ which are not accessible in $Z$-decay.

\bigskip
\bigskip

\noindent $\bullet$ CHARGED CURRENT NEUTRINO PROCESS\\

\bigskip

\noindent {\it Mu-neutrino electron scattering}

\bigskip
\nobreak
The reaction $\nu_\mu e^- \rightarrow \mu^- \nu_e$ the simplest
four-fermion $CC$ scattering process.
We shall follow the convention of ascribing the four-momentum
$p(f)$ to the fermion $f$ and write the four-momentum transfer to the
target as $q$.  Define
$$
s \equiv \left[p(\nu_\mu) + p(e)\right]^2, ~~~t \equiv q^2 \equiv -Q^2 =
\left[p(\nu_\mu) - p(\mu)\right]^2,
$$
so that in the physical region $0 < Q^2 < s$.  We also introduce the
inelasticity variable
$$
y \equiv p(e) \cdot \left[p(\nu_\mu) - p(\mu)\right]/p(e) \cdot p(\mu) =
\left[(E_{\rm initial} - E_{\rm final})/E_{\rm initial}\right]_{\rm
lab}
$$
lying in the kinematic range $0 < y < 1$.  It is also convenient to
introduce the $W$-propagator function (we ignore the $W$-width)
$$
R_W (Q) \equiv {M^4_W \over (M^2_W + Q^2)^2}
$$
which is nearly unity when $Q^2 \ll M^2_W$.

\vspace{3in}

For CM energies much larger than the electron mass $(s \gg m^2_e)$, the
lowest order differential cross section is given by
$$
{d\sigma^{\nu_\mu e}_{CC} \over dy} = {G^2_F s \over \pi} R_W (Q).
\eqno (5)
$$
The lack of $y$-dependence in the RHS of (5) is a hallmark of the left-chiral
$V-A$ interaction of the $W$.  In contrast, a right-chiral $V+A$
interaction would have yielded an extra $(1-y)^2$ factor.  In fact, by
comparing (5) with experiment, an upper limit $|g_R/g_L| < 0.0039$ has
been obtained [6] on the ratio of the magnitudes of a right-chiral and a
left-chiral semiweak coupling.  In the largely available kinematic range
$m^2_e \ll Q^2$, $s \ll M^2_W$, (5) can be integrated to yield the total
cross section formula
$$
\sigma^{\nu_\mu e}_{CC} = {G^2_F s \over \pi}.
\eqno (6)
$$
The linear rise of the total cross section with $s$ in the relevant
kinematic range is experimentally well-established and the value of the
coefficient has been verified at the 5\% level [6].

\bigskip

\noindent {\it Mu-neutrino isoscalar nucleon deep inelastic CC scattering}\\

\nobreak
The reactions $(\nu_\mu,\bar\nu_\mu)N \rightarrow \mu^\mp X$ are studied
using a heavy approximately isoscalar nuclear target and an external muon
identifier.  As before

\vspace*{3in}

$$
0 < Q^2 \equiv -\left[p(\nu_\mu) - p(\mu)\right]^2 < s \equiv
\left[p(\nu_\mu) + p(N)\right]^2
$$
and
$$
0 < y \equiv {p(N) \cdot \left[p(\nu_\mu) - p(\mu)\right] \over p(N) \cdot
p(\nu_\mu)} = \left[(E_{\rm initial} - E_{\rm final})/E_{\rm
initial}\right]_{\rm lab} < 1.
$$
The deep inelastic region corresponds to $s,Q^2$ being $\gg M^2_N$ with
$Q^2/s$ fixed.  Formally, this is reached by the limit $Q^2 \rightarrow
\infty$ with $Q^2/2p(N) \cdot [p(\nu_\mu) - p(\mu)] \equiv \omega$ fixed.

In this scaling limit ($\omega$ being the scaling variable) the asymptotic
freedom property of Quantum Chromodynamics (QCD) allows a parton model
description [7].  According to this, the deep inelastic scattering cross
section can be described (Fig. 6) as an incoherent sum of elementary scattering
processes of the neutrino from quark and antiquark partons folded by the
parton distribution function $\int dx q_i (x,Q^2)$ for the $i^{\rm th}$
type of parton.  Here $x$ is the longitudinal fraction of
the nucleon momentum carried by the parton in an infinite
momentum frame.  This variable $x$ is constrained to equal $\omega$
through a $\delta(x-\omega)$ which arises as a factor in the elementary
cross section and kills the $x$-integration.  The relevant elementary
processes of $\nu_\mu$ and $\bar\nu_\mu$ scattering from quarks and
antiquarks are shown in Fig. 7.

\vspace{3in}

\vspace*{3in}

Let $u,d$ refer to the up, down type of quark and $i(= 1,2,3)$
to the generation.  The quark and antiquark distribution functions
are reasonably well-known [8] from QCD studies.  Isospin invariance
implies
$$
\left[\matrix{u_i(x,Q^2) \cr di(x,Q^2)}\right]_p = \left[\matrix{di(x,Q^2)
\cr u_i(x,Q^2)}\right]_n
$$
and similarly for antiquarks, with $p(n)$ referring to a target proton
(neutron).  Utilizing these relations, one can write all quark
distributions $q_i(x,Q^2)$ and antiquark distributions $\bar q_i(x,Q^2)$
with respect to a proton only.

Define
$$
\begin{array}{l}
{\cal Q} (Q^2) \equiv \sum_i \int^1_0 dx q_i(x,Q^2), \\[2mm]
\overline{\cal Q} (Q^2) \equiv \sum_i \int^1_0 dx \bar q_i (x,Q^2)
\end{array}
\eqno (7)
$$
and also consider notionally an isoscalar nucleon $N \equiv {1\over2}
(p+n)$ as the target.  Now the QCD Parton Model expressions for the
differential cross sections of our proceses (ignoring inter-generation
transitions) can be written as
$$
{d\sigma^{\nu_\mu N}_{CC} \over dy} = {G^2_F s \over 2\pi} R_W (Q^2)
\left[{\cal Q} (Q^2) + (1-y)^2 \overline{\cal Q} (Q^2)\right],
\eqno (8a)
$$
$$
{d\sigma^{\bar\nu_\mu N}_{CC} \over dy} = {G^2_F s \over 2\pi} R_W (Q^2)
\left[{\cal Q} (Q^2) (1-y)^2 + \overline{\cal Q} (Q^2)\right].
\eqno (8b)
$$
For the large kinematic range $M^2_N \ll s$, $|t| \ll M^2_W$, (8)
integrates to the total cross section formulae
$$
\sigma^{\nu_\mu N}_{CC} = {G^2_F s \over 6\pi} \left[3\langle {\cal Q}
\rangle + \langle \overline{\cal Q}\rangle\right],
\eqno (9a)
$$
$$
\sigma^{\bar\nu_\mu N}_{CC} = {G^2_F s \over 6\pi} \left[\langle {\cal
Q}\rangle + 3\langle \overline{\cal Q}\rangle\right],
\eqno (9b)
$$
where angular brackets have been put on ${\cal Q},\overline{\cal Q}$ since
a $Q^2$-averaging takes place along with the $y$-intergration.  Once
again, the linear $s$-dependence is well-tested [9].

\bigskip

\noindent {\it Neutrinos from ep scattering}

\bigskip
\nobreak
Electron neutrinos and antineutrinos can be produced through the $CC$
processes
$$
\begin{array}{l}
e^- p \rightarrow \nu_e X,\\[2mm]
e^+ p \rightarrow \bar\nu_e X,
\end{array}
$$
as will be studied shortly in HERA which has 820 GeV protons
colliding with 30 GeV $e^-$ or $e^+$.  The corresponding differential
cross sections are
$$
{d\sigma^{e^-p \rightarrow \nu_e X}_{CC} \over dy} = {G^2_F s \over 2\pi}
R_W (Q) \left[{\cal U} (Q^2) + \overline{\cal D}(Q^2) (1-y)^2\right],
\eqno (10a)
$$
$$
{d\sigma^{e^+p \rightarrow \bar\nu_e X}_{CC} \over dy} = {G^2_F s \over
2\pi} R_W (Q) \left[{\cal D} (Q^2) (1-y)^2 + \overline{\cal U}(Q^2)\right].
\eqno (10b)
$$
In (10)
$$
{\cal U} (Q^2) = \sum_i \int^1_0 dx u_i (x,Q^2),
\eqno (11a)
$$
$$
{\cal D} (Q^2) = \sum_i \int^1_0 dx di (x,Q^2),
\eqno (11b)
$$
$$
\overline{\cal U} (Q^2) = \sum_i \int^1_0 dx \bar u_i (x,Q^2),
\eqno (11c)
$$
$$
\overline{\cal D} (Q^2) = \sum_i \int^1_0 dx \bar d_i (x,Q^2).
\eqno (11d)
$$
Integrated cross sections $\sim 50$ pb are expected [10].

One can also have muon neutrinos and antineutrinos produced, along with
charged dileptons, through the reactions
$$
e^\pm p \longrightarrow {e^+ \mu^- \bar\nu_\mu X \atop e^- \mu^+
\nu_\mu X}.
$$
The corresponding Feynman diagrams are given below.  Integrated cross
sections $\sim 5 \times 10^{-2}$ pb are expected [10] at HERA.

\vspace*{5.8in}

\bigskip
\bigskip

\noindent $\bullet$ NEUTRAL CURRENT NEUTRINO PROCESSES \\

\nobreak
\noindent {\it Mu-neutrino electron scattering}\\

The elastic scattering reaction $(\nu_\mu,\bar\nu_\mu)e \rightarrow
(\nu_\mu,\bar\nu_\mu)e$ ~is one of the sim-

\newpage
\noindent plest $NC$-induced four fermion
processes.

\vspace*{3in}

\noindent One can define $s,t$ and $Q^2$ in analogy with the $\nu_\mu e
\rightarrow \mu^- \nu_e$ case and take the inelasticity variable to be
$$
y \equiv q \cdot p(e) \left[p_{\rm initial} (\nu_\mu) \cdot
p(e)\right]^{-1} = \left[(E_{\rm initial} - E_{\rm final})/E_{\rm
initial}\right]_{\rm lab}
$$
Furthermore, in the zero-width approximation, the $Z$ propagator function is
$$
R_Z (Q) = {M^4_Z \over (Q^2 + M^2_Z)^2}.
$$
We shall also find it convenient to write $x_W$ for the sine squared of
the Weinberg angle $\sin^2 \theta_W = 1 - M^2_W/M^2_Z$.

The lowest order differential cross sections can now be written, in the
limit when $s \gg m^2_e$, as
$$
{d\sigma^{\nu_\mu e}_{NC} \over dy} = {G^2_F s \over \pi} R_Z (Q)
\left[g^2_L + g^2_R (1-y)^2\right],
\eqno (12a)
$$
$$
{d\sigma^{\bar\nu_\mu e}_{NC} \over dy} = {G^2_F s \over \pi} R_Z (Q)
\left[g^2_{Le} (1-y)^2 + g^2_{Re}\right],
\eqno (12b)
$$
with
$$
g_{Le} = - {1\over2} + x_W,
\eqno (13a)
$$
$$
g_{Re} = x_W.
\eqno (13b)
$$
Here $g_{Le}$ and $g_{Re}$ respectively define the left-chiral and
right-chiral $Ze\bar e$ couplings.  In (12a), the latter contribution has a
$(1-y)^2$ coefficient, but not the former.  One can compare this situation
with the corresponding charged current case.  For the scattering of the
$\bar\nu_\mu$ off an electron, the role of the $(1-y)^2$ factor gets
reversed; it now multiplies the left-chiral rather than the right-chiral
contribution.

As in the $CC$ case, for $m^2_e \ll Q^2,s \ll M^2_Z$, (12) can be
integrated to yield total cross sections
$$
\sigma^{\nu_\mu e}_{NC} = {G^2_F s \over 4\pi} \left(1 - x_W + {16 \over
3} x^2_W\right) \simeq 1.6 \times 10^{-2} (E_{\rm initial}/10~{\rm
GeV})fb,
$$
$$
\sigma^{\bar\nu_\mu e}_{NC} = {G^2_F s \over 4\pi} \left({1\over3} -
{4\over3} x_W + {16 \over 3} x^2_W\right) \simeq 1.3 \times 10^{-2}
(E_{\rm initial}/10~{\rm GeV})fb.
$$
An important ratio from which $x_W$ and hence $\theta_W$ -- the mixing
angle between $SU(2)_L$ and $U(1)_Y$ -- can be directly determined is
$$
{\sigma^{\nu_\mu e}_{NC} \over \sigma^{\bar\nu_\mu e}_{NC}} = {1 - 4 x_W +
{16\over3} x^2_W \over {1\over3} - {4\over3} x_W + {16 \over 3} x^2_W}.
\eqno (14)
$$
In extracting $x_W$ from $\nu_\mu$ and $\bar\nu_\mu$ elastic scattering
from the same target of atomic electrons, one has to carefully monitor the
relative fluxes of the $\nu_\mu,\bar\nu_\mu$ beam components.  Of course,
many systematic errors cancel out in the ratio.  Finally, complete 1-loop
radiative corrections have to be accounted for; fortunately, this has now
been done.  Employing all these steps the CHARM II experiment [11] has
determined
$$
x_W = 0.233 \pm 0.012 \pm 0.008,
\eqno (15)
$$
where the second error is statistical and the third is the estimated
systematic error.

\bigskip

\noindent {\it Mu-neutrino isoscalar nucleon deep inelastic NC scattering}\\

\nobreak
\vspace*{4in}

We employ the same notation as in the corresponding $CC$ induced deep
inelastic neutrino scattering.  Thus we can write $NC$ differential cross
sections analogous to (8) as
$$
\begin{array}{l}
d\sigma^{\nu_\mu N}_{NC}/dy = (2\pi)^{-1} G^2_F s ~R_Z (Q)
\Bigg[(g^2_{Lu} + g^2_{Ld}) \{{\cal Q} (Q^2) + \overline{\cal Q} (Q^2)
(1-y)^2\} \\[2mm]
{}~~~~~~~~~~~~~~~~~~~~~~~~~~~~~~~~~~~~~~~~~+ (g^2_{Ru} + g^2_{Rd}) \{(1-y)^2
{\cal Q} (Q^2) + \overline{\cal Q} (Q^2)\}\Bigg],
\end{array}
\eqno (16a)
$$
$$
\begin{array}{l}
d\sigma^{\bar\nu_\mu N}_{NC}/dy = (2\pi)^{-1} G^2_F s ~R_Z (Q)
\Bigg[(g^2_{Lu} + g^2_{Ld}) \{(1-y)^2 {\cal Q} (Q^2) + \overline{\cal Q}
(Q^2)\} \\[2mm]
{}~~~~~~~~~~~~~~~~~~~~~~~~~~~~~~~~~~~~~~~~~~+ (g^2_{Ru} + g^2_{Rd}) \{{\cal
Q} (Q^2) + \overline{\cal Q} (Q^2) (1-y)^2\}\Bigg]
\end{array}
\eqno (16b)
$$
and
$$
g^2_{Lu} + g^2_{Ld} = {1\over2} - x_W + {5\over9} x^2_W,
\eqno (17a)
$$
$$
g^2_{Ru} + g^2_{Rd} = {5\over9} x^2_W.
\eqno (17b)
$$
We have defined $g_{Lq}$ and $g_{Rq}$ $(q = u,d)$ in analogy with $g_{Le}$
and $g_{Re}$.  For the neutrino case, the expression multiplying the
left-chiral couplings has $(1-y)^2$, 1 as the coefficient of the antiquark,
quark distribution while that multiplying the right-chiral couplings has 1,
$(1-y)^2$ as the coefficient of the antiquark, quark distribution.  The
situation is exactly reversed for the antineutrino case.  Once again, for
$M^2_N \ll Q^2$, $s \ll M^2_Z$, (16) can be trivially integrated to give
total cross sections
$$
\sigma^{\nu_\mu N}_{NC} = {G^2_F s \over 2\pi} \left\{\left({1\over2} -
x_W + {20 \over 27} x^2_W\right) \langle {\cal Q}\rangle + {1\over3}
\left({1\over2} - x_W + {20 \over 9} x^2_W\right) \langle \overline{\cal
Q}\rangle \right\},
\eqno (18a)
$$
$$
\sigma^{\bar\nu_\mu N}_{NC} = {G^2_F s \over 2\pi} \left\{{1\over3}
\left({1\over2} - x_W + {20\over9} x^2_W\right) \langle {\cal Q}\rangle +
\left({1\over2} - x_W + {20 \over 27} x^2_W\right) \langle \overline{\cal
Q}\rangle\right\}.
\eqno (18b)
$$

The relations (16) are very useful -- especially with respect to
extracting $x_W$ and hence $\theta_W$.  First, note a useful equality due
to Llewellyn Smith [12]:
$$
{d\sigma^{(\nu_\mu,\bar\nu_\mu)N}_{NC} \over dy} = \left({1\over2} - x_W +
{5\over9} x^2_W\right) {d\sigma^{(\nu_\mu,\bar\nu_\mu)N}_{CC} \over dy} +
{5\over9} x^4_W {d\sigma^{(\bar\nu_\mu,\nu_\mu)N}_{CC} \over dy}.
\eqno (19)
$$
The advantage of (19) is that it does not involve the quark and antiquark
distributions in a nucleon.  The disadvantage is that the relative fluxes
of the $\nu_\mu$ and $\bar\nu_\mu$ beams need to be calculated accurately.
In fact, many ratios, called Paschos-Wolfenstein [13] ratios, are independent
of quark and antiquark distributions since $\{{\cal Q} (Q^2) \pm
\overline{\cal Q} (Q^2)\} \{1 \pm (1-y)^2\}$ can be factored out by
taking the sum or difference of (16a) and (16b).  Thus
$$
\left({d\sigma^{\nu_\mu N}_{NC} \over dy} + {d\sigma^{\bar\nu_\mu N}_{NC}
\over dy}\right) \left({d\sigma^{\nu_\mu N}_{CC} \over dy} +
{d\sigma^{\bar\nu_\mu N}_{CC} \over dy}\right)^{-1} = {1\over2} - x_W +
{10\over9} x^2_W
\eqno (20a)
$$
and
$$
\left({d\sigma^{\nu_\mu N}_{NC} \over dy} - {d\sigma^{\bar\nu_\mu N}_{NC}
\over dy}\right) \left({d\sigma^{\nu_\mu N}_{CC} \over dy} -
{d\sigma^{\bar\nu_\mu N}_{CC} \over dy}\right)^{-1} = {1\over2} - x_W.
\eqno (20b)
$$

If we define $R_\nu \equiv \sigma^{\nu_\mu N}_{NC}/\sigma^{\nu_\mu
N}_{CC}$, $R_{\bar\nu} \equiv \sigma^{\bar\nu_\mu N}_{NC}/
\sigma^{\bar\nu_\mu N}_{CC}$ and $r \equiv \sigma^{\bar\nu_\mu N}_{CC}/
\sigma^{\nu_\mu N}_{CC}$, $R_\nu$ and $R_{\bar\nu}$ become
flux-independent quantities while $r$ is sensitive to the knowledge of the
relative fluxes of the $\nu_\mu$ and $\bar\nu_\mu$ beams.  (18) can now be
recast as
$$
R_\nu = {1\over2} - x_W + {5\over9} x^2_W (1 + r),
\eqno (21a)
$$
$$
R_{\bar\nu} = {1\over2} - x_W + {5\over9} x^2_W \left(1 +
{1\over r}\right),
\eqno (21b)
$$
This relation (22) is totally flux-independent and provides a clean means
of extracting $x_W$.  Using such relations, the CHARM group determined [14]
$$
x_W = 0.233 \pm 0.003 \pm 0.005.
$$

\bigskip

\noindent $\bullet$ PROPERTIES OF VERY HEAVY NEUTRINOS \\

\bigskip
\nobreak

\noindent {\it Preliminaries}\\

\bigskip
\nobreak

A massive neutrino (generically described by a field $N$) can be one of
two types : Dirac or Majorana, depending on whether it is different from
its antiparticle or identical to it.  The antiparticle is described by the
charge-conjugated field $N^C$.  A Dirac neutrino field $(N /\!\!\!\!\!=
N^C)$ has four distinct chiral components : $N_L,N_R,N_L^{~C} = N^C_{~R},
N^{~C}_R = N^C_{~L}$.  It has a mass term in the Lagrangian given by
$$
{\cal L}^D_m = -M^D_N (\bar N_L N_R + \bar N_R N_L).
\eqno (22)
$$
A Majorana neutrino, in contrast, has $N = N^C$ and hence only two
distinct chiral components : $N_L = N^C_{~L} = N^{~C}_R$ and $N_R = N^C_{~R} =
N^{~C}_L$.  Though it can be described by a 2-component formalism, we will
find it convenient to use the same formalism as in the Dirac case with the
proviso $N = N^C$.  The Majorana mass term in the Lagrangian, which
violates lepton number, can be written as
$$
{\cal L}^M_m = -{1\over2} M^M_N \bar N^C N + h.c. = -{1\over2} M^M_N N^T
C^{-1} N + h.c.,
\eqno (23)
$$
where $C$ is the charge conjugation matrix with
$$
\begin{array}{l}
N^C = (\bar N C)^T,\\[2mm]
\gamma^0 C^\star \gamma^0 = C^{-1}.
\end{array}
$$
We shall consider very heavy neutrinos [15] with mass $\gsim$ tens of GeV
that can be wither Dirac or Majorana type.

\bigskip

\noindent {\it Theoretical motivation}\\

\bigskip
\nobreak
There exist a number of scenarios in which such very heavy neutrinos are
expected to occur.  We outline a few.

\bigskip

\noindent 1. {\sl Fourth generation model} -- This has been proposed by
Hill and Paschos [5] and has a heavy charged lepton $\ell$ and a
neutrino $N$ making a fourth replica of the existing three generations.
Thus one has a left-chiral doublet and two right-chiral singlets
$$
\left(\matrix{N \cr \ell^-}\right)_L, ~~~\ell^-_R, ~~~N_R.
$$
Existing LEP constraints from $Z$-decay simply require that $M_N >
\displaystyle {1\over2} M_Z$.  One need not have $N_R$ but then would be
forced to invoke the lepton-number violating Majorana mass term (23).

\bigskip

\noindent 2. {\sl Left-right symmetric model} -- The simplest such model
employs the gauge group $SU(2)_L \times SU(2)_R \times U(1)_{B-L}$ and
contains one left-chiral and one right-chiral lepton doublet per generation:
$$
\left(\matrix{\nu \cr e}\right)_L,~~~\left(\matrix{N \cr e}\right)_R.
$$
(We have distinguished here between $\nu$ and $N$ since we are leading up
to two physically different neutrinos per generation: one very light and
one very heavy).  One can have a seesaw mass term (with $M \gg m$)
$$
{\cal L}_m = -{1\over2} m \bar\nu_L N_R - {1\over2} M N^T_R C^{-1} N_R +
h.c. = -{1\over2} (\bar\nu_L \bar N^{~C}_R) \left(\matrix{0 & m \cr m &
M}\right) \left(\matrix{\nu_L^{~C} \cr N_R}\right) + h.c.
\eqno (24)
$$
The eigenvalues of the seesaw mass-matrix, $\simeq M$ and $\displaystyle
{m^2 \over M}$ (the negative sign of the latter being absorbed by a chirality
transformation $\psi \rightarrow \gamma_5 \psi$) describe a very heavy
and a very light physical neutrino.  References to discussions of their
phenomenology can be found in [16].

The very heavy neutrino, described in the above two scenarios, can be
searched for in LEP 200 via the reaction $e^+e^- \rightarrow Z^\star
\rightarrow N\bar N$ provided its mass is less than a 100 GeV.  It can, in
principle, be looked for by exploiting its mixing with $\nu_e$ via the
production mode $ep \rightarrow NX$ at HERA but the estimated cross
sections [16] look impossibly small.  However, there are two other
scenarios as discussed below.

\bigskip

\noindent 3. {\sl Pure singlet model}  -- In this case there is an extra
right-chiral singlet heavy neutrino $N_R$ for each generation.  Thus, for
the first, one has
$$
\left(\matrix{\nu \cr e}\right)_L, ~C_R, ~N_R.
$$
$N_R$ can have a large Majorana mass and it is possible to arrange a
seesaw mass-matrix between $\nu_L$ and $N_R$.  The main importance of this
type of a model is that [17] $N$ can be produced singly in $e^+e^-$ or
$ep$ collision, as detailed later.

\bigskip

\noindent 4. {\sl $E_6$-based models} -- The grand unifying gauge group
$E_6$ is very popular among model builders starting with the $E_8 \times
E_8$ superstring.  The matter fields in an $E_6$ GUT, arising from the
topological breakdown of one of the $E_8$'s, belong to the
{\bf 27} dimensional representation of $E_6$.  This can
accommodate three extra heavy neutrinos [8], their masses depending on
various symmetry-breaking scales in the breakdown chain $E_6 \rightarrow
SU(3)_C \times SU(2)_L \times U(1)_Y$.  Two of these neutrinos are
$SU(2)_L \times U(1)_Y$ singlets.  The third, transforming as part of a
vectorial doublet with respect to weak isospin, directly couples to $W$
and $Z$.  The others too can mix with the usual very light neutrinos
$(\nu_\ell,\ell = e,\mu,\tau)$ and develop couplings to the weak bosons.

\bigskip

\noindent {\it Heavy neutrino couplings and decays}\\

\bigskip
\nobreak
There are far too many model-independent possibilities in the pattern of
couplings of such very heavy neutrinos to the known elementary particles.
We try to adopt a generic approach following [17].  Let us assume that, on
account of mixing with $\nu_\ell$, $N$ develops a charged current coupling
to $W\ell$ and neutral current couplings to $NZ$ as well as $\nu_\ell Z$
-- as shown below.  Here

\vspace*{3in}

\noindent $\xi$ is a small sessaw mixing factor (hopefully $\gsim
10^{-3}$) and $V_{N\ell}$ is a Kobayashi-Maskawa type matrix element.  (It
may be noted that the model of [5] cannot be covered by this since there
is no $ZN\bar\nu_\ell$ vertex there and no mixing factor in the $ZN\bar N$
coupling).  We are also obliged to choose $M_N > M_Z$, otherwise -- for
$\xi > 10^{-3}$ -- the decays $Z \rightarrow M\bar\nu_\ell,\bar N\nu_\ell$
would have already been seen at LEP.

We come to the decays of $N$.  First, consider the case when $N$ is a
Dirac particle.  Now, for the charged current mode
$$
\Gamma (N \rightarrow \ell^+ W^+) = \Gamma(\bar N \rightarrow \ell^+ W^-)
= {|\xi V_{N\ell}|^2 \over 8\sqrt{2} \pi} {G_F \over M^3_N} (M^2_N +
2M^2_W) (M^2_N - M^2_W)^2.
\eqno (25)
$$
Contrariwise, for the neutral current mode
$$
\Gamma(N \rightarrow \nu_\ell Z) = \Gamma(\bar N \rightarrow \bar\nu_\ell
Z) = {|\xi|^2 \over 8\sqrt{2}\pi} {G_F \over M^3_N} (M^2_N + 2M^2_Z)
(M^2_N - M^2_Z)^2.
\eqno (26)
$$
The equality of the $N$ and $\bar N$ partial widths in (25) and (26)
follows from $CP$-invariance.  The charged current mode is less dominant
than the neutral current one since the latter does not have the small
$|V_{N\ell}|^2$ factor.
For a Majorana heavy neutrino, $N$ and its antiparticle are identical and
one simply has $\Gamma(N \rightarrow \ell^- W^+) = \Gamma(N \rightarrow
\ell^+ W^-)$ and $\Gamma(N \rightarrow \nu_\ell Z) = \Gamma(N \rightarrow
\bar\nu_\ell Z)$, with the corresponding expressions still given by (25)
and (26) respectively.  Thus the lifetime of $N$ gets halved as compared
with a Dirac N.  In either case, for $M_N \gg M_{W,Z}$ and $\xi \gsim
10^{-3}$, the mean free path is $\ll$ cms.  Thus, if produced in the
laboratory, such an $N$ will decay within the detector.

Though the neutral current induced decay is the dominant mode, the charged
current mediated one $(N \rightarrow \ell W)$ can provide the cleanest
signals for detection.  The $W$ can decay into two jets so that $\ell(2j)$
is the detectable final state configuration.  For the Dirac case and with
a pair-produced $N\bar N$, one would have the hard signal $\ell^+
\ell^{^{\prime \atop -}} (4j)$ where $\ell$ and $\ell^\prime$ need not be
the same.  (Of course, one would have to tackle the severe background from
the semileptonic decays of top-antitop pairs).  If a pair of Majorana $N$'s
gets prodeuced, we can have three possibilities : $\ell^+ \ell^{^{\prime
\atop -}} (4j), ~\ell^+ \ell^{^{\prime \atop +}} (4j)$ and $\ell^-
\ell^{^{\prime \atop -}} (4j)$.  While these are characteristic signals,
one cannot exclude a very heavy neutrino in the relevant mass-range simply
by failing to observe them.  This is because certain models allow [8] the
dominant decay $N \rightarrow \nu J$ where $J$ is a very light
pseudo-Goldstone boson like a Majoron.  The experimental unobservability of
this decay channel would make it harder to discover $N$.

\bigskip

\noindent {\it Production mechanisms}\\

\bigskip
\nobreak
First, we take up the production of single $N$'s.  In an $e^+e^-$ collider
this can be done through the processes $e^+e^- \rightarrow
N\bar\nu_\ell,\bar N\nu_\ell$.  These go via the Feynman diagrams of Fig. 13.
Asymptotically, for a large $CM$ energy
$\sqrt{s}$, the cross section is approximately
$\pi^{-1} G^2_F M^2_W |V_{\ell N} \xi|^2$.
Note that for a Majorana $N$ there are only three diagrams since (c) and
(d) become one and the same.  In any event, the $Z$-mediated part
contributes only about 2\% of the total cross section.  This it is a good
approximation to take only the $W$-mediated part.  Typically, a fraction
of a picobarn is expected at LEP 200 as shown in Fig. 14.

\newpage

\vspace*{10in}

\newpage

\noindent In Fig. 14 the production cross section [17] has been plotted against
$\sqrt{s}$ as
well as against the heavy neutrino mass $M_N$.  Coming to
electroproduction $e^-p \rightarrow NX$, say at HERA, the cross
section is
shown against $M_N$ for various values of $\sqrt{s}$.  Of course, the
signal (Fig. 15)
will depend on whether the $N$ decays into $\ell W$ or $\nu_\ell
Z$, but $\ell^- (2j) /\!\!\!\!E_T$, $\ell^+\ell^- /\!\!\!\!E_T$, and
$\ell^+\ell^{^{\prime \atop -}} /\!\!\!\!E_T$ are possible signal
configurations.  It has recently been suggested [18] that the process
$e^-\gamma \rightarrow W^-N$ (Fig. 16) would be a viable production
mechanism in a 1
TeV $e^+e^-$ linear collider with a cross section $\sim |\xi V_{N\ell}|^2$
pb.

\vspace{4in}

Next, we come to pair-production.  This can be attained through the $ZN\bar
N$ coupling which is perhaps less model-dependent than the $ZN\bar \nu$
one.  We put a generic mixing factor $\chi$ to cover the cases where $N$
is an $SU(2)_L \times U(1)_Y$ singlet.  (For a regular fourth generation
heavy neutrino, $\chi$ is unity).  The cross section for $e^+e^-
\rightarrow Z^\star \rightarrow N\bar N$ (Fig. 17) can be calculated [8] to be

\newpage

\vspace*{3in}

$$
\sigma = {G^2_F s \over 24\pi} \left(1 - {4M^2_N \over s}\right)^{1/2}
\left(1 - {M^2_N \over s}\right) R_Z (Q) |\chi|^2 (1 - 4x_W + 8x^2_W).
\eqno (27)
$$
In the high-energy limit when $s \gg M^2_Z$, the RHS of (27) goes as $2.5
\times 10^{-2} |\chi|^2$ (pb/s in TeV$^2$) which is about $0.6 |\chi|^2$
pb for $\sqrt{s} = 200$ GeV at LEP 200.  Similar considerations hold for
the Drell-Yan type of production process $q\bar q \rightarrow Z^\star
\rightarrow N\bar N$ in a hadron collider.  One problem with the cross
section of (27) is the rapid fall off of the factor $R_Z (Q)$ at large $s$
which drastically reduces the cross section at supercollider energies
$\gsim 1$ TeV.

We shall discuss an alternative mechanism of heavy neutrino\\
pair-production via the fusion of two gluons [19] which is relevant to
$pp$ supercolliders.  As shown in Fig. 18, two gluons from the colliding
protons can
go via a quark loop into an off-shell $Z$ boson which converts
into an $N\bar N$ pair.  The heavy neutrinos, in turn, decay into $\ell
(2j)$ and $\ell'(2j)$, say so that the signal configuration is $\ell\ell'
(4j)$.  For a Majorana pair, one can have like sign dileptons which with
four jets make an almost unique signal for this process.  ~~In the case
of a~ Dirac pair and a ~signal ~configuration ~of

\newpage

\vspace*{5in}

\noindent  $\ell^+\ell^{^{\prime
\atop -}}
(4j)$, the background from $t\bar t$ pair-production and subsequent
semileptonic decays of $t,\bar t$ would be overwhelming.  But now one can
hook on to the leptonic decay of one of the $W$'s and search for the
signal $\ell^+ \ell^{^{\prime \atop -}} \ell^{^{\prime\prime \atop +}} (2j)
/\!\!\!\!E_T$ or $\ell^+ \ell^{^{\prime \atop -}} \ell^{^{\prime\prime
\atop -}} (2j) /\!\!\!\!E_T$.  There are several characteristic features
of this mechanism:

\bigskip

$\bullet$ The triangular loop has a nonzero contribution only from the
axial part of the $Z$-coupling, the vector part vanishing on account of
Furry's theorem.

$\bullet$ The contributing part is an anomaly graph proportional, not only
to the third component of the weak isospin of the fermion circulating ~in
{}~the

\newpage

\noindent triangle, but also -- through the divergence of the axial current --
to its mass.  Thus the mass difference $|M_U - M_D|$ between the up-type
and down-type quarks of the heaviest generation comes in the numerator of
the dominant part of the amplitude.

$\bullet$ By Yang's theorem, the amplitude is nonzero only because of the
off-shell nature of the $Z$.  The consequent $Q^2 - M^2_Z$ in the
numerator cancels the denominator from the propagator making the cross
section only weakly dependent on $s$.

\newpage

It turns out that, with the three known generations of quarks, the cross
section from the above mechanism will not be measurable at supercollider
energies.  However, if there is a fourth generation of quarks with $|m_U -
m_D| \sim 100$ GeV, then one can get quite decent cross sections (fb to
tens of fb) both at SSC and LHC.  Fig. 19 shows these cross
sections as a function of $M_N$ both for $\sqrt{s} = 16$ TeV (LHC) and
$\sqrt{s} = 40$ TeV (SSC).  The bands are generated by variations in $m_V$
(from 400 GeV to 1200 GeV) with $|m_V - m_D|$ kept $\sim 100$ GeV.

It has recently [20] been realized that the Higgs-mediated $gg \rightarrow
H^\star \rightarrow N\bar N$ cross section is enhanced relative to the
$Z$-mediated one to which it adds incoherently on-account of the
difference in the $s$-channel angular momentum.  One can actually have
both scalar Higgs $H$ and pseudoscalar Higgs $P$ exchanges (in models with
more than one
doublet) which also add incoherently.  Choudhury et al have, in fact,
demonstrated (taking $m_{\rm top} = 160$ GeV) that the cross section at
SSC energies in either case is expected to be quite large (pb to
sub-pb range) even with just the three known generations of quarks.  This
is displayed in Figs. 20 below.

\newpage

\noindent $\bullet$ CONCLUDING REMARKS\\

\bigskip
\nobreak
In covering the salient features of accelerator-based neutrino physics, we
hope to have brought forth in a reasonably persuasive way that the subject
is alive and well.  Right now two major high energy neutrino beam
experiments are in progress, being undertaken by the CCFR and CHARM (2)
groups.  The level of precision in the data (with respect to both
statistics and systematics) continues to become more impressive day by
day.  Those aspects of the Standard Model which are probed by high energy
neutrino beams are very well tested.  It is also possible to perform
neutrino oscillation experiments with accelerator-generated neutrino
beams.  Indeed, such experiments have been designed at Fermilab and CERN.
But I have chosen to leave those topics to speakers focusing on neutrino
oscillations.

Apart from scattering experiments with neutrino beams, final state
neutrinos have been and are being produced and studied.  LEP 1 has made a
major contribution in this direction.  HERA as well as LEP 200 and the
forthcoming $pp$ supercolliders LHC and SSC do and will provide
opportunities to search for new very heavy neutrinos ($\sim 10^2$ GeV in
mass).  The best bet seems to be Higgs-mediated pair-production in $pp$
supercolliders.

\bigskip

\noindent {\it Acknowledgements}

\bigskip
\nobreak
I am very grateful to Jihn E. Kim for inviting me to lecture in this
delightful summer school and also for giving me the opportunity to see
Korea.  I thank K.V.L. Sarma for several helpful discussions.

\newpage

\begin{center}
\large {\bf REFERENCES}\\
\end{center}

\begin{enumerate}

\item[{[1]}] F. Reines and C.L. Cowan, Phys. Rev. {\bf 113} (1957) 273.

\item[{[2]}] C.H. Llewellyn Smith, Phys. Rep. {\bf 3C} (1972) no. 5.

\item[{[3]}] Review of Particle Properties, Phys. Rev. {\bf D45}, Part 2
(June 1992).

\item[{[4]}] A. Gurtu, Proc. X DAE Symp. {\it High Energy Physics}, Pramana
J. Phys. (suppl.), in press.

\item[{[5]}] C.T. Hill and E. Paschos, Phys. Lett. {\bf B241} (1990) 96.

\item[{[6]}] S.R. Mishra, Nucl. Phys. B (Proc. Suppl.) {\bf 19} (1991) 193.

\item[{[7]}] R. Feynman, {\it Photon-Hadron Interactions} (Benjamin, 1972).
P. Roy, {\it Theory of Lepton-Hadron Processes at High Energies}
(Clarendon Press, Oxford, 1975).  F. Close, {\it Introduction to Quarks
and Partons} (Academic Press, London, 1979).

\item[{[8]}] V. Barger and R.J.N. Phillips, {\it Collider Physics}
(Addison-Wesley, corrected edition, 1988).

\item[{[9]}] S.R. Mishra and F. Sciulli, Ann. Rev. Nucl. Part. Sci. {\bf 39}
(1989) 259.

\item[{[10]}] M.M.J.F. Janssen, Nationaal IUnstitut voor Kernfysica en
Hodge-Energifysica Report NIKHEP-H/91-07.

\item[{[11]}] D. Geiregat et al., Phys. Lett. {\bf 232B} (1989) 539.

\item[{[12]}] C.H. Llewellyn Smith, Nucl. Phys. {\bf B228} (1983) 205.

\item[{[13]}] E. Paschos and L. Wolfenstein, Phys. Rev. {\bf D7} (1973) 91.

\item[{[14]}] P.G. Reutens et al., Phys. Lett. {\bf 152B} (1985) 404.

\item[{[15]}] M. Gronau, C.N. Leung and J.L. Rosner, Phys. Rev. {\bf D29}
(1984) 2539.  H.E. Haler and M.R. Reno, Phys. Rev. {\bf D31} (1986) 2372.

\item[{[16]}] G. Altarelli, B. Mele and R. Ruckl, CERN-ECFA Workshop {\bf 2}
(1984) 549 (QCD 183:E2:1984).

\item[{[17]}] W. Buchm\"uller and C. Greub, Phys. Lett. {\bf B256} (1991)
465; Nucl. Phys. {\bf B381} (1992) 109.

\item[{[18]}] M.C. Gonzalez-Garcia, O.J. \'Eboli, F. Halzen and S.F. Novaes,
Phys. Lett. {\bf B280} (1992) 313.

\item[{[19]}] D.A. Dicus and P. Roy, Phys. Rev. {\bf D44} (1991) 1593.

\item[{[20]}] A. Datta and A. Pilaftsis, Phys. Lett. {\bf B278} (1992) 162;
D. Choudhury, R.M. Godbole and P. Roy, Phys. Lett. {\bf B308} (1993) 394.

\end{enumerate}
\end{document}